\documentclass[namedreferences, sophpaper]{solarphysics}
\pdfoutput=1

\usepackage[hyperref, optionalrh]{spr-sola-addons}
\def\UrlFont{\sf}
\usepackage[utf8]{inputenc}
\usepackage[T1]{fontenc}
\usepackage{graphicx}
\usepackage{color}
\usepackage{natbib}
\usepackage{makecell}
\usepackage{savesym}
\savesymbol{tablenum}
\usepackage{siunitx}
\restoresymbol{SIX}{tablenum}
\usepackage{xspace}
\usepackage{float}
\usepackage{apjfonts}

\hypersetup{
  pdftitle={Hinode/EIS Roll Angle},
  pdfauthor={G. Pelouze et al.},
  pdfkeywords={Hinode, EIS; Instrumental Effects; Instrumentation and Data Management; Spectrum, ultraviolet}
  }

\def\UrlFont{\sf}
\def\FileFont{\sf}
\let\originalhref\href
\renewcommand\href[2]{\originalhref{#1}{\UrlFont#2}}

\newcommand\narrowfigwidth{.7\textwidth}

\newcommand{\etal}{\textit{et al.}\xspace}
\newcommand{\ie}{\textit{i.e.}\xspace}
\newcommand{\eg}{\textit{e.g.}\xspace}

\usepackage[dvipsnames]{xcolor}
\newcommand\corr[1]{#1}

\newcommand\ionelem[2]{#1~{\scshape{#2}}\xspace}
\newcommand\ionline[3]{#1~{\scshape{#2}}~\SI{#3}{\angstrom}\xspace}

\begin{document}
\begin{article}
\begin{opening}

\title{Comprehensive Determination of the \textit{Hinode}/EIS Roll Angle}
\runningtitle{\textit{Hinode}/EIS Roll Angle}
\runningauthor{G. Pelouze \etal}

\author[addressref={ias},corref,email={gabriel.pelouze@ias.u-psud.fr}]{\inits{G.}\fnm{Gabriel}~\lnm{Pelouze}\orcid{0000-0002-0397-2214}}%
\author[addressref={ias}]{\inits{F.}\fnm{Frédéric}~\lnm{Auchère}\orcid{0000-0003-0972-7022}}%
\author[addressref={ias}]{\inits{K.}\fnm{Karine}~\lnm{Bocchialini}\orcid{0000-0001-9426-8558}}%
\author[addressref={mssl}]{\inits{L.}\fnm{Louise}~\lnm{Harra}\orcid{0000-0001-9457-6200}}%
\author[addressref={mssl}]{\inits{D.}\fnm{Deborah}~\lnm{Baker}\orcid{0000-0002-0665-2355}}%
\author[addressref={ssd-nrl}]{\inits{H.}\fnm{Harry P.}~\lnm{Warren}\orcid{0000-0001-6102-6851}}%
\author[addressref={cs-gmu}]{\inits{D.}\fnm{David H.}~\lnm{Brooks}\orcid{0000-0002-2189-9313}}%
\author[addressref={spacs-gmu}]{\inits{J.}\fnm{John T.}~\lnm{Mariska}\orcid{0000-0002-5463-2375}}%

\address[id={ias}]{Institut d’Astrophysique Spatiale, CNRS, Univ. Paris-Sud, Université Paris-Saclay, Bât. 121, 91405 Orsay cedex, France}
\address[id={mssl}]{Mullard Space Science Laboratory, University College London, Holmbury, St. Mary, Dorking, Surrey, KT22 9XF, UK}
\address[id={ssd-nrl}]{Space Science Division, Naval Research Laboratory, Washington, DC 20375, USA}
\address[id={cs-gmu}]{College of Science, George Mason University, 4400 University Drive, Fairfax, VA 22030 USA}
\address[id={spacs-gmu}]{Department of Physics and Astronomy, George Mason University, 4400 University Drive, Fairfax, VA 22030, USA}

\begin{abstract}
% Context
% Aims
We present a new coalignment method for the \textit{EUV Imaging Spectrometer} (EIS) on board the \textit{Hinode} spacecraft. In addition to the pointing offset and spacecraft jitter, this method determines the roll angle of the instrument, which has never been systematically measured, and is therefore usually not corrected.
% Methods
The optimal pointing for EIS is computed by maximizing the cross-correlations of the \ionline{Fe}{xii}{195.119} line with images from the \SI{193}{\angstrom} band of the \textit{Atmospheric Imaging Assembly} (AIA) on board the \textit{Solar Dynamics Observatory} (SDO).
By coaligning 3336 rasters with high signal-to-noise ratio, we estimate the rotation angle between EIS and AIA and explore the distribution of its values.
% Results
We report an average value of \ang{-0.387 \pm .007;;}.
% Conclusions
We also provide a software implementation of this method that can be used to coalign any EIS raster.
\end{abstract}


\end{opening}

\clearpage

\section{Introduction}
\label{s:intro}

To analyze data from the \textit{Extreme-ultraviolet (EUV) Imaging Spectrometer} \citep[EIS:][]{CulhaneEtAl2007} on board \textit{Hinode} \citep{KosugiEtAl2007}, it is required to accurately correct the pointing of the instrument. This is usually done by registering (\ie finding the geometrical transform between two images) EIS rasters with images from the \textit{Atmospheric Imaging Assembly} \citep[AIA:][]{LemenEtAl2012} on board the \textit{Solar Dynamics Observatory} \citep[SDO:][]{PesnellEtAl2012}, or from the \textit{Extreme-ultraviolet Imaging Telescope} \citep[EIT:][]{DelaboudiniereEtAl1995} on board the \textit{Solar and Heliospheric Observatory} \citep[SOHO:][]{DomingoEtAl1995}. 

Registering EIS images with the reference instrument requires knowledge of two translations (the pointing offsets along the \textit{X}- and \textit{Y}-axis), a scaling factor (the ratio of the plate scales), and a rotation (the roll-angle difference in the plane of the sky). The roll angle is the most difficult parameter to determine. 

The roll angles of the \textit{X-Ray Telescope} \citep[XRT:][]{GolubEtAl2007} and the \textit{Solar Optical Telescope} \citep[SOT:][]{TsunetaEtAl2008}, both on board \textit{Hinode}, have been determined using transits of Mercury by \citet{ShimizuEtAl2007}.
The temporal evolution of the XRT roll angle was later measured by \citet{YoshimuraMcKenzie2015}, who used correlations with AIA and the \textit{Helioseismic and Magnetic Imager} \citep[HMI:][]{ScherrerEtAl2012} on board SDO. The authors find that this rotation angle changes periodically over one year with an amplitude of about \ang{0.2;;}.
The roll angle of EIS, however, has never been determined, and is therefore not systematically accounted for.
Indeed, very few studies report taking into account a rotation when coaligning EIS data \citep[see, \eg,][who report doing so]{BrooksEtAl2012}.

Furthermore, the spacecraft jitter randomly changes \corr{the spacecraft attitude} by a few arcsec on all three axes and at each slit position during a raster scan \citep{ShimizuEtAl2007}.
This has a significant effect on the pointing offset, which can change by several pixels as a result. Therefore each slit position has to be coaligned independently from one another, along the \textit{X}- and \textit{Y}-directions.
However, the effect of the jitter on the roll angle (\textit{Z}-axis) is negligible, because it rotates the field of view (FOV) by \corr{a few arcsec} in the plane of the sky \corr{(the same angle as around the \textit{X}- and \textit{Y}-axes)}. Over the largest EIS field of view (\ang{;;512}), a rotation \corr{of \ang{;;10}} would shift the observed structures by less than \ang{;;0.02} (or \num{0.02}~pixels) at the edge of the FOV. Because of this, we can search for an overall rotation angle common to all slit positions.

In \autoref{s:method}, we present a new method to register EIS rasters with SDO/AIA images, that corrects the instrument roll and spacecraft jitter. In \autoref{s:results}, we apply this method to get an accurate estimation of the roll angle between the spectrometer EIS and the imager AIA. In \autoref{s:discussion}, we investigate the temporal dependency of this roll angle, and conclude that it is consistent with the findings of \citet{YoshimuraMcKenzie2015} for XRT. Finally, we summarize our results and provide software to coalign EIS rasters with AIA in \autoref{s:conclusion}.

\section{A New Method to Register EIS Rasters with AIA}
\label{s:method}

\subsection{Overview}
\label{s:method_overview}

We determine the pointing for an EIS raster by searching for the maximum cross-correlation between an \ionline{Fe}{xii}{195.119} intensity map and a synthetic raster built from AIA~\SI{193}{\angstrom} images. 
This synthetic raster simulates what would be seen by AIA if it acquired images by scanning each column at a time, as it is the case with EIS rasters.
For efficiency, the \ionelem{Fe}{xii} map is obtained by summing intensities between \num{194.969} and \SI{195.269}{\angstrom} from the level~1 EIS raster, which has been prepared with the routine {\FileFont eis\_prep.pro} from SolarSoft \citep{FreelandHandy2012}. We verified that identical registration results are obtained when the intensities are computed using {\FileFont eis\_auto\_fit.pro}, which fits Gaussians to the \ionelem{Fe}{xii} \num{195.119} and \SI{195.179}{\angstrom} lines.
The synthetic raster is obtained from a cube of level~1 AIA~\SI{193}{\angstrom} images with a cadence of one minute, from which intensities are derived at the EIS spatial and temporal positions using bilinear interpolation. The synthetic raster is then degraded to the resolution of EIS by convolving it with a gaussian PSF of \ang{;;3} FWHM (\citealp{DelZannaEtAl2011_warm_loops}; \corr{\citealp{YoungEtAl2013}}).
\corr{In order for the AIA data to be as close as possible to what would be observed by EIS, we use different synthetic rasters when computing the cross-correlation map, which are generated for each sampled value of offset and rotation.}
Using synthetic rasters instead of a single AIA image is necessary because structures on the Sun may change significantly during the acquisition of the EIS raster, which can take up to several hours.
\corr{
We use the plate scale value of \ang{;;1}~$\mathrm{pixel}^{-1}$  from the EIS headers. 
Although this value slightly differs from the \ang{;;1.002 \pm 0.016}~$\mathrm{pixel}^{-1}$ reported by \citet{Hara2008}, comparison between aligned EIS and AIA images shows no significant deformation of the structures that could be caused by an incorrect plate scale.}

Searching for the global maximum cross-correlation in one run would require excessive computation time because the parameter space to explore is very large.
Therefore we perform the correction in three sequential steps to save time: 1.~determine and correct the average translation; 2.~determine and correct the roll angle; 3.~correct the jitter by coaligning each slit position independently. 
At step~1, we search for a translation that can be as large as the raster field of view.
At step~2, we simultaneously search for rolls around the center of the field of view with angles between \num{-3} and \ang{3;;}, along with a smaller translation (between \num{-10} and \ang{;;10} along the \textit{X}-axis, and \num{-5} and \ang{;;5} along the \textit{Y}-axis).
At step~3 we search for translations of each slit position (\ie each column of the raster), between \num{-20} and \ang{;;20} along both axes.
These search limits were chosen by computing the cross-correlation over a wider range of offsets for about a hundred of rasters, and looking at the distributions of the maximum position.

\subsection{Performance}

\begin{figure}[ht]
\centerline{\includegraphics[width=\narrowfigwidth]{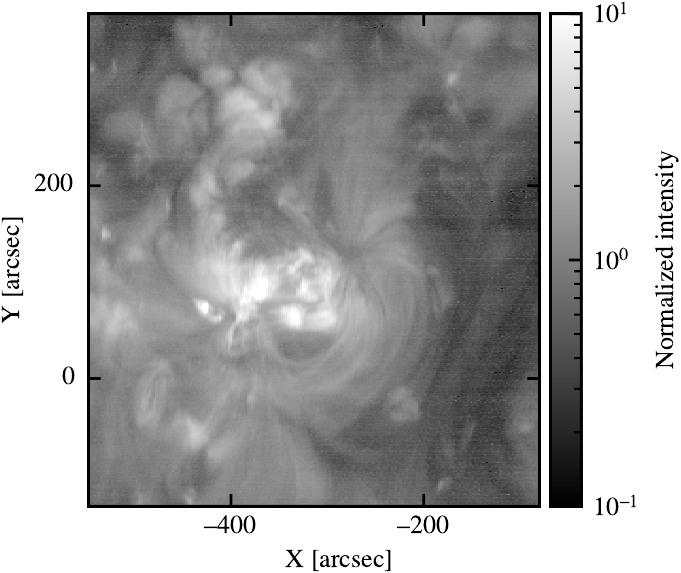}}
\caption{Map of the \ionline{Fe}{xii}{195.119} line emission for raster
  {\FileFont eis\_l0\_20140810\_042212}.
  The intensity is normalized to its standard deviation over the field of view,
  and the axes are labelled with the original EIS pointing.
    }
\label{fig:example_4steps_intensity_map}
\end{figure}

\begin{figure}[ht]
\centerline{\includegraphics[width=\narrowfigwidth,clip]{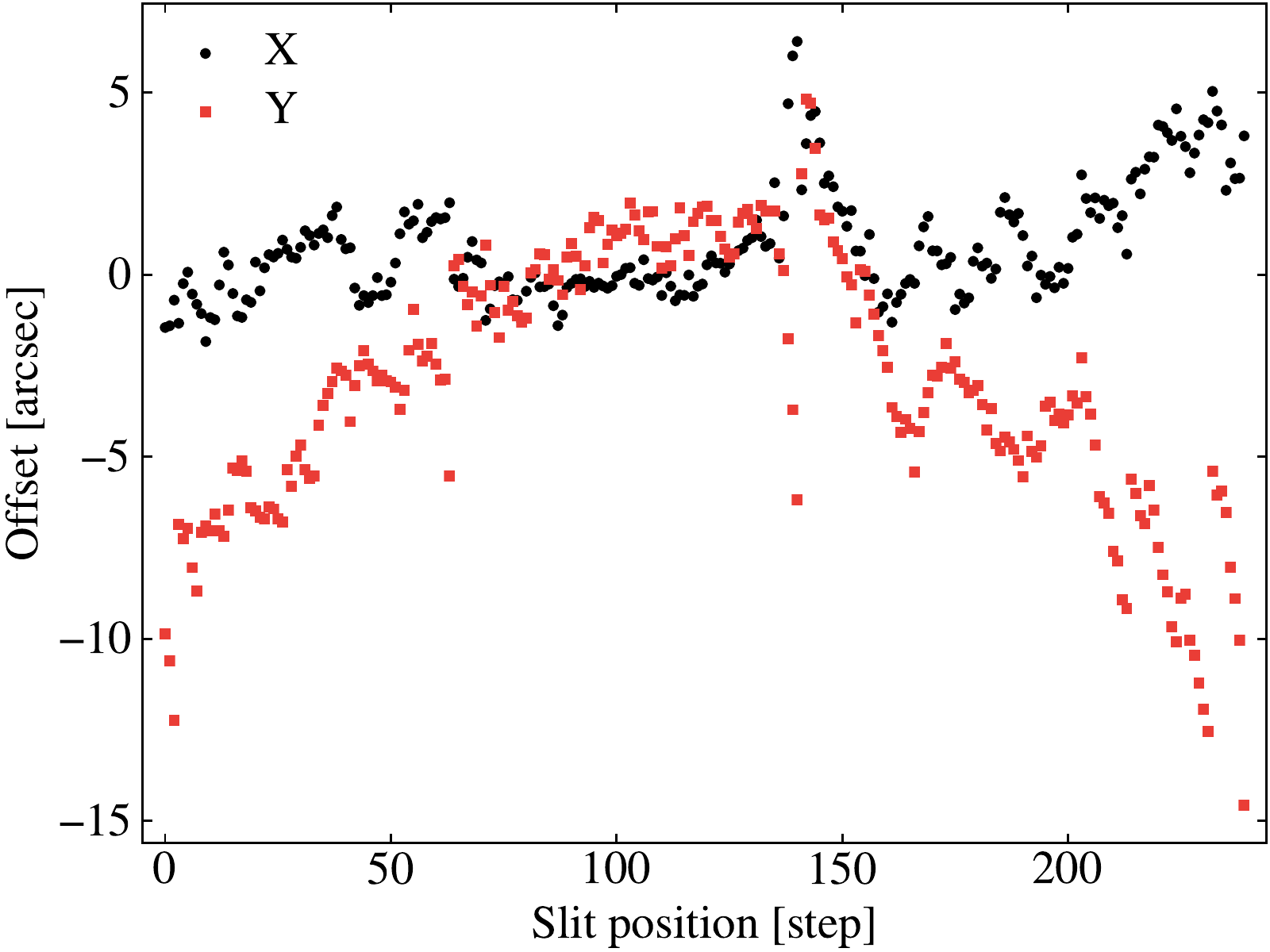}}
\caption{
  Slit offsets obtained when correcting the jitter for raster
  {\FileFont eis\_l0\_20140810\_042212}.
  Black dots represent offsets along the \textit{X}-direction,
  while red squares represent offsets along the \textit{Y}-direction.
  }
\label{fig:slit_align}
\end{figure}

For rasters with sufficient SNR (exposure times greater than 15~seconds with the \ang{;;1} slit for on-disk observations), our method can efficiently correct the visible deformation of the structures due to the instrument rotation and satellite jitter. 
We present registration results for raster {\FileFont eis\_l0\_20140810\_042212}, which corresponds to the observation of the active region NOAA~12135 on 10~August~2014 starting from 04:22:12~UT. This raster features an exposure time of 15~seconds at each position of the \ang{;;1} slit, a scan step of \ang{;;2}, and a field of view of $\ang{;;480}\times\ang{;;512}$. The field of view of this raster is shown in \autoref{fig:example_4steps_intensity_map}.
The registration gives an offset of $(\ang{;;17.0}, \ang{;;10.6})$, a roll angle of \ang{-0.78;;}, and the slit offsets that are plotted in \autoref{fig:slit_align}.
For this raster, the dispersion of the slit offset values is of about \ang{;;2} to \ang{;;3}, and the offset along the \textit{Y}-axis shows additional large scale variations of about \ang{;;15}. Other rasters have a similar dispersion, but do not share any large-scale variation pattern.
\corr{The dispersion and absolute values of the slit offset don’t appear to be correlated with the \textit{Hinode} eclipse season, during which the Sun is periodically eclipsed by the upper Earth atmosphere, causing larger average pointing offsets \citep{EISSWN20, YoshimuraMcKenzie2015}.}
In order to obtain the corrected coordinates, these offsets are added to the original EIS coordinates, and the image is rotated clockwise.

\begin{figure}[ht]
\centerline{\includegraphics[width=\textwidth]{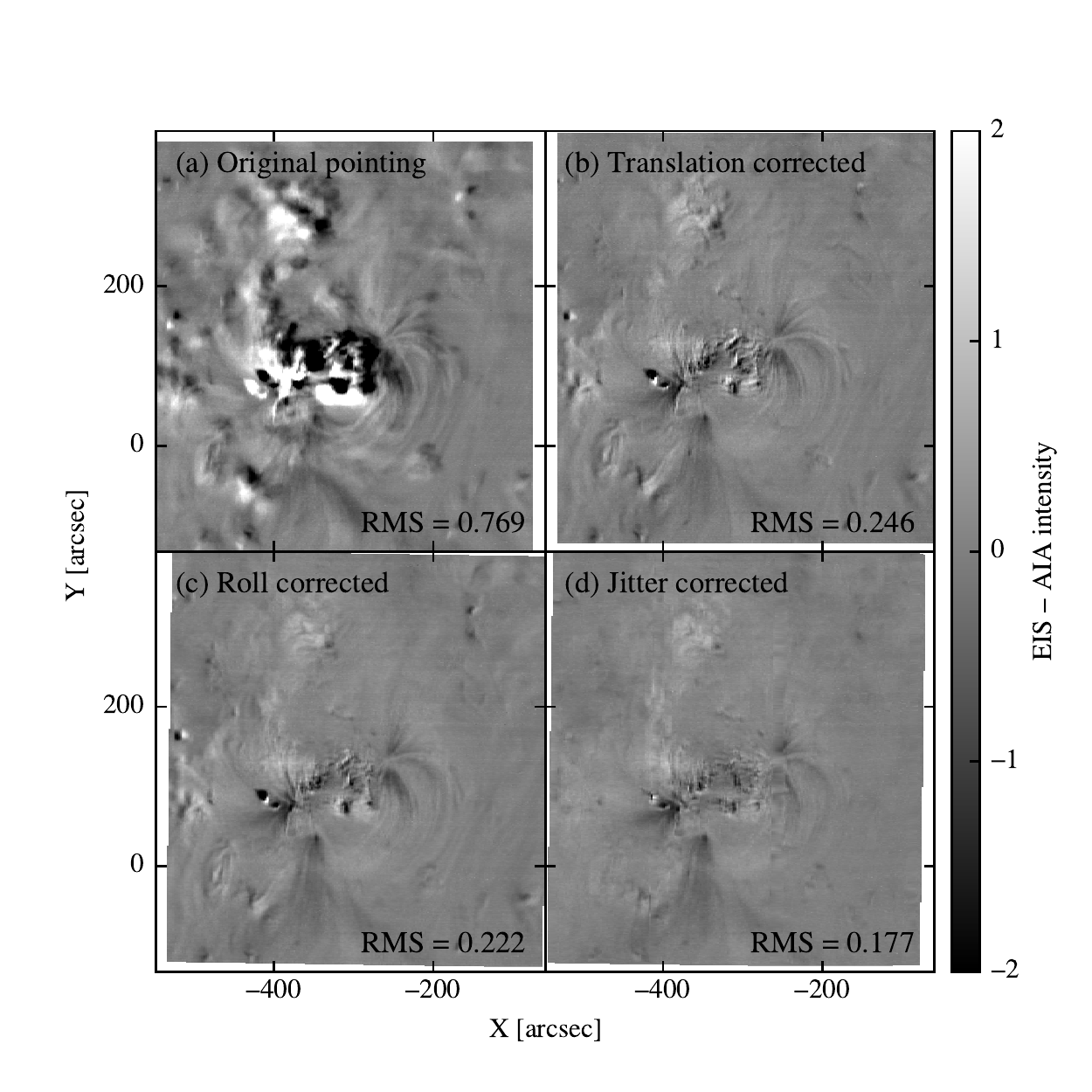}}
\caption{Registration result for raster {\FileFont eis\_l0\_20140810\_042212}. Each
  panel contains a map of the difference between normalized EIS and AIA
  intensities at each step of the registration:
  \textit{(a)}~original EIS pointing,
  \textit{(b)}~average translation corrected,
  \textit{(c)}~roll corrected, and
  \textit{(d)}~jitter corrected.
  The root mean square shown on each panels quantifies the difference between
  structures observed in EIS and AIA.
  An animated version of this figure is available as an electronic supplementary material,
  where the EIS and AIA intensity maps are shown alternately;
  this allows us to better visualize the rotation and deformations of the structures.
  }
\label{fig:example_4steps}
\end{figure}

In \autoref{fig:example_4steps}, we show the difference between aligned EIS and AIA intensity maps, which have been normalized to their respective standard deviations.
An animated version of this figure is also available as an electronic supplementary material, which allows us to better visualize the rotation and deformations in the field of view by showing the EIS and AIA intensity maps alternately.
The visible variation between structures observed in both images is quantified by the root mean square (RMS) of this normalized intensity difference.
The original data have a RMS of 0.769, which decreases to 0.177 after the registration is performed.
In addition to alignment errors, the residuals are affected by differences in solar structures when observed in the EIS~\ionline{Fe}{xii}{195.119} line, or the AIA~\SI{193}{\angstrom} channel.

\begin{figure}[ht]
\centerline{\includegraphics[width=\textwidth]{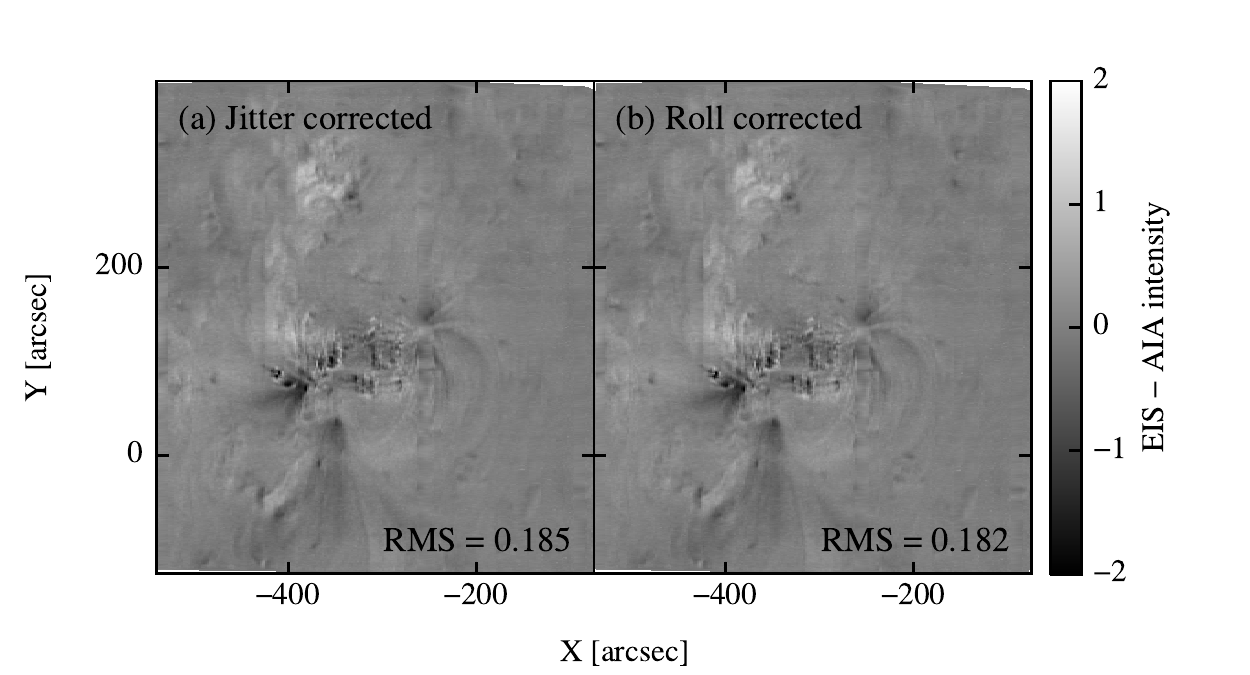}}
\caption{Registration result for raster {\FileFont eis\_l0\_20140810\_042212}, swapping steps 2 and 3: 
  \textit{(a)} jitter corrected, and
  \textit{(b)} roll corrected.
  (See \autoref{fig:example_4steps} for the reference alignment.)
  An animated version of this figure is available as an electronic supplementary material.
}
\label{fig:example_4steps_rev}
\end{figure}

We performed additional testing to validate the method.
First, we correct the jitter before correcting the roll angle (\ie swapping registration steps~2 and~3). As shown in \autoref{fig:example_4steps_rev}, this achieves worse results with a rotation still visible in the animated version, and a final RMS of only 0.182.
We also apply a second roll correction after correcting the jitter. This does not improve the pointing and results in the same final RMS of 0.177.
Finally, we verify whether the measured roll angle may be caused by tilts between the internal components of EIS.
The \textit{slit tilt} is a known angle between the slits and the EIS detectors, which slants spectral lines on the detector \citep{EISSWN4}. On the short wavelength (SW) detector, lines are rotated by \ang{0.03;;} with the \ang{;;1} slit, and by \ang{0.3;;} with the \ang{;;2} slit. While this can significantly modify measured velocities, it should have no influence on the pointing.
However, the pointing would be affected by an angle between the slits and the tilt axis of the mirror, which defines the direction along which a raster is scanned. This would result in a sheared image, meaning that each slit position appears to be rotated by a given angle relatively to the scanning direction. We aligned raster {\FileFont eis\_l0\_20140810\_042212} by replacing the search for a roll angle (step 2) with the search for a shear transform. Because this could not correct the visible rotation of the field of view, we rule out the presence of a rotation between the slits and the tilt-mirror mechanism as the source of the observed roll.

From these tests, we conclude that three registration steps are required in order to accurately register EIS maps with AIA, and that the best results are achieved when they are applied in the following order : translation, rotation, and jitter.

\section{Results: EIS Roll Angle}
\label{s:results}

To accurately measure the roll angle between EIS and \textit{Hinode} we correlate a large number of rasters with AIA. We query the EIS database (\href{http://sdc.uio.no}{sdc.uio.no}) for rasters recorded between 13~May~2010 and 4~September~2018 that match the following criteria:
the center of the raster is on-disk,
the field of view is wider than \ang{;;200},
the raster uses either the \ang{;;1} or the \ang{;;2} slit,
and the exposure time is longer than 15~seconds. 
The query returns 3856~rasters that we process with the method described in \autoref{s:method}. 

\begin{figure}[ht]
\centerline{\includegraphics[width=\narrowfigwidth,clip=]{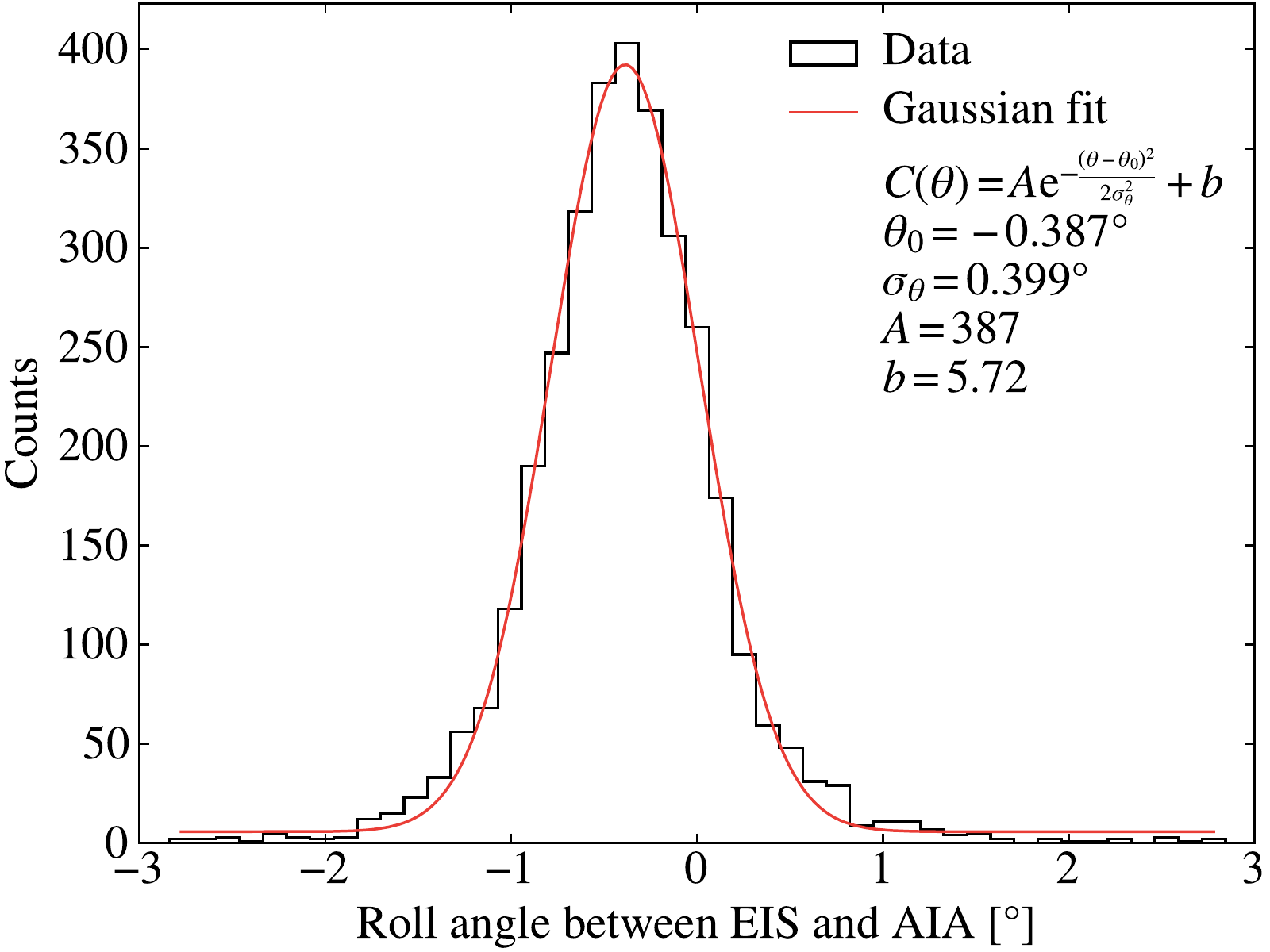}}
\caption{
  Histogram of the roll angle between EIS and AIA measured on 3336 rasters between 2010 and 2018.
  The red line is a Gaussian that fits the histogram, with a center at \ang{-0.387;;}, and a standard deviation of \ang{0.399;;}.
  }
\label{fig:hist_rot_fit}
\end{figure}

Most of the rasters are successfully registered, which results in 3707 measurements of the roll angle between EIS and AIA.
\corr{80 of the failed measurements were caused by bad initial EIS pointing (file headers indicate a raster center far outside the disk), 48 by missing or corrupted AIA data, and 21 by errors during the EIS data preparation or too many missing pixels in the resulting raster.}
In order to discard bad registrations, we remove \SI{10}{\percent} of the alignment results for which the RMS of the EIS and AIA intensity difference is the largest.
The histogram of the $N = 3336$ remaining roll-angle values is shown in \autoref{fig:hist_rot_fit}.
This distribution can be approximated by a Gaussian centered at $\theta_0 = \ang{-0.387;;}$, and with a standard deviation of $\sigma_\theta = \ang{0.399;;}$. 
The uncertainty on $\theta_0$ is given by $\sigma_\theta / \sqrt{N}$. 
Therefore, we estimate the average roll angle between EIS and AIA to be $\ang{-.387 \pm 0.007;;}$.
The dispersion $\sigma_\theta$ can be due to a combination of measurement errors, and/or true variations of the roll angle.
\corr{For this reason, the uncertainty on the roll angle for an arbitrary raster is larger than \ang{0.007;;}.}
Thus to coalign an EIS raster with an AIA image, the EIS raster must be rotated by \ang{-0.387;;}, \ie a clockwise rotation.

\section{Discussion}
\label{s:discussion}

\begin{figure}[ht]
\centerline{\includegraphics[width=\narrowfigwidth,clip=]{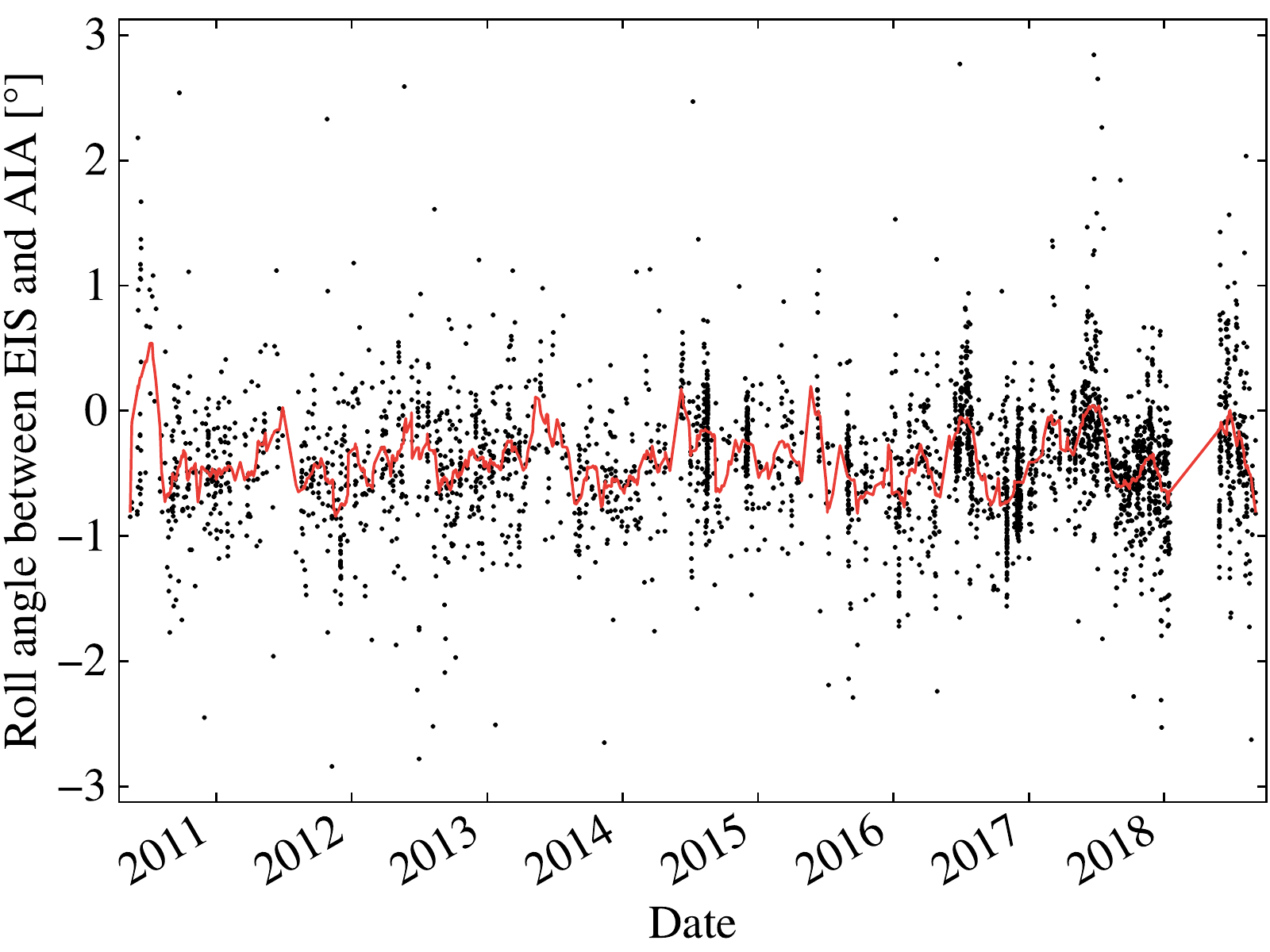}}
\caption{%
  Time series of the roll angle between EIS and AIA measured on 3336 rasters between 2010 and 2018.
  Individual measurements are plotted as black dots, and a 20-day boxcar running average is shown as a red line.
}
\label{fig:angles_time_series}
\end{figure}

\begin{figure}[ht]
\centerline{\includegraphics[width=\narrowfigwidth,clip=]{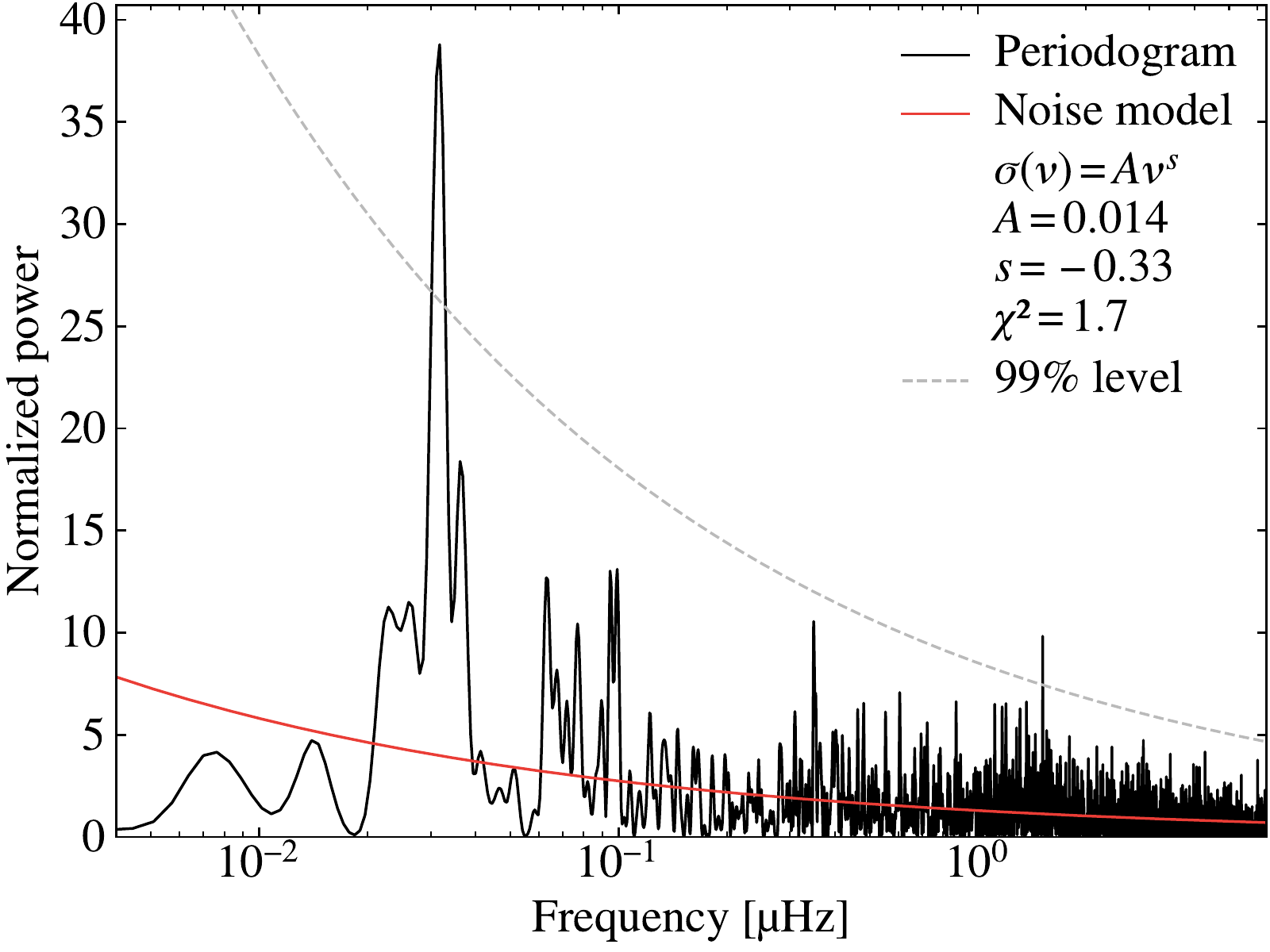}}
\caption{%
  Lomb--Scargle periodogram of the time series of the roll angle between EIS and AIA measured on 3336 rasters between 2010 and 2018.
  The power-law noise model is shown as a solid red line, and the \SI{99}{\percent} confidence level is shown as a gray dashed line.
}
\label{fig:periodogram}
\end{figure}

In order to understand the distribution of roll-angle values, we investigate how they depend on several parameters, such as exposure time, field-of-view size, orbit phase, raster duration, scan step, and slit width. We find that none of these parameters significantly affect the roll angle.
We also search for structures in the time series of roll-angle values, which is shown in \autoref{fig:angles_time_series}. While there appears to be no secular evolution of the average values, we also search for periodic variations. 
To that end, we estimate the power spectral density (PSD) of the roll-angle time series using a Lomb--Scargle periodogram \citep{Lomb1976, Scargle1982} shown in \autoref{fig:periodogram}.
To model the noise, we fit the PSD with a power law $\sigma(\nu) = A \nu^s$, which yields $A = 0.013$, and $s = -0.35$. 
The probability that at least one peak has a power greater than $m \sigma(\nu)$ is $P(m) = 1 - (1 - \mathrm{e}^{-m})^{N_i}$, where $N_i$ is the number of independent frequencies \citep{Scargle1982, GabrielEtAl2002, AuchereEtAl2016a}. 
However, estimating the number of independent frequencies for unevenly spaced data is difficult. In our case, the rasters are sometimes closely clumped in time, which can significantly reduce the number of independent frequencies (relatively to the case of evenly spaced measurements), and therefore lower confidence levels \citep{HorneBaliunas1986}.
% The waiting times between two measurements have a very wide distribution, ranging from a few minutes to several weeks, which indicates an important deviation even sampling. 
We estimate an upper bound for the confidence levels by assuming that the samples are regularly spaced, which implies that $N_i = N$.
The periodogram reveals two peaks above the \SI{99}{\percent} confidence level: one with a period of one year, and the other with a period of \SI{7.6}{days}.

The first peak is consistent with the results of \citet{YoshimuraMcKenzie2015}, who find that the roll angle between XRT and AIA or HMI changes periodically over one year, with an amplitude of about \ang{0.2;;} and an average of \ang[retain-explicit-plus]{+0.34;;}.
They also find that the roll angle is largest at the end of July each year, and smallest at the end of January.
While the one-year variation is not visible directly in our raw time series, applying a 20-day boxcar running average, shown in \autoref{fig:angles_time_series}, reveals that the roll angle between EIS and AIA is also at its maximum around July.
The fact that the roll angles between EIS and AIA (this study), and between XRT and AIA or HMI \citep{YoshimuraMcKenzie2015} display similar variations but have different averages suggests that the constant part comes from alignment offsets of the instruments with respect to \textit{Hinode}, while the variations come from the behaviour of the spacecrafts (\textit{Hinode} and/or SDO), regardless of the instruments.

Obtaining true solar coordinates for EIS should require no additional step once it is registered with AIA, because we use the pointing information present in the AIA level~1 image headers, which are corrected for plate scale, telescope coalignment, and roll relatively to the solar North \citep{LemenEtAl2012, BoernerEtAl2012}.

\section{Conclusion}
\label{s:conclusion}

We developed a new method to register EIS rasters with SDO/AIA, which can determine and correct the pointing offsets and the roll angle between the two instruments, as well as the \textit{Hinode} jitter.
By applying this method to 3336 rasters recorded between May 2010 and September 2018, we were able to measure the average roll angle between EIS and AIA to a value of \ang{-0.387 \pm .007;;}.
Such a rotation shifts the structures by \ang{;;3.5} over \ang{;;512}, which is the size of large EIS rasters.
\corr{We also found evidence for a time dependency of this roll angle, which results in a standard deviation of \ang{.399;;}. For this reason, we recommend that the roll angle is determined for each raster whenever possible.}

The implementation of our registration method is provided as a Python package, which can be found at \href{https://github.com/gpelouze/eis_pointing}{github.com/gpelouze/eis\_pointing}. This tool can be used to coalign EIS rasters with AIA images when precise spatial information is required. While correcting the jitter requires high SNR (typically 15~second exposure time with the \ang{;;1} slit for on-disk observations), the determination of the pointing offset and of the roll angle should work on most rasters.

\begin{acks}
\corr{The authors would like to thank the anonymous referee for contributing to the improvement of this paper.}
\textit{Hinode} is a Japanese mission developed and launched by ISAS/JAXA, with NAOJ as domestic partner and NASA and STFC (UK) as international partners. It is operated by these agencies in co-operation with ESA and NSC (Norway).
AIA data are courtesy of NASA/SDO and the AIA science team.
This work used data provided by the MEDOC data and operations centre (CNES/CNRS/Univ. Paris-Sud), \href{http://medoc.ias.u-psud.fr}{medoc.ias.u-psud.fr}.
D.B. is funded under STFC consolidated grant number ST/N000722/1.
The work of D.H.B., \corr{J.T.M., and H.P.W.} was performed under contract with the Naval Research Laboratory and was funded by the NASA \textit{Hinode} program.
\textit{Software:}
Astropy \citep{Astropy2013, Astropy2018},
SolarSoft \citep{FreelandHandy2012}.
\end{acks}

\paragraph*{\footnotesize Disclosure of Potential Conflicts of Interest}
The authors declare that they have no conflicts of interest.

\bibliographystyle{spr-mp-sola}
\bibliography{eis_roll}

\begin{thebibliography}{26}
% BibTex style file: spr-mp-sola.bst (nameyear), 2015-03-09
\ifx\bisbn     \undefined \def\bisbn  #1{ISBN #1}\fi
\ifx\binits    \undefined \def\binits#1{#1}\fi
\ifx\bauthor   \undefined \def\bauthor#1{#1}\fi
\ifx\batitle   \undefined \def\batitle#1{#1}\fi
\ifx\bjtitle   \undefined \def\bjtitle#1{\textit{#1}}\fi
\ifx\bvolume   \undefined \def\bvolume#1{\textbf{#1}}\fi
\ifx\byear     \undefined \def\byear#1{#1}\fi
\ifx\bissue    \undefined \def\bissue#1{#1}\fi
\ifx\bfpage    \undefined \def\bfpage#1{#1}\fi
\ifx\blpage    \undefined \def\blpage #1{#1}\fi
\ifx\burl      \undefined \def\burl#1{\textsf{#1}}\fi
\ifx\href      \undefined \def\href#1#2{\textsf{#2}}\fi
\ifx\betal     \undefined \def\betal{\textit{et al.}}\fi
\ifx\bctitle   \undefined \def\bctitle#1{#1}\fi
\ifx\beditor   \undefined \def\beditor#1{#1}\fi
\ifx\bbtitle   \undefined \def\bbtitle#1{\textit{#1}}\fi
\ifx\bedition  \undefined \def\bedition#1{#1}\fi
\ifx\bseriesno \undefined \def\bseriesno#1{\textbf{#1}}\fi
\ifx\blocation \undefined \def\blocation#1{#1}\fi
\ifx\bsertitle \undefined \def\bsertitle#1{\textit{#1}}\fi
\ifx\bsnm      \undefined \def\bsnm#1{#1}\fi
\ifx\bsuffix   \undefined \def\bsuffix#1{#1}\fi
\ifx\bparticle \undefined \def\bparticle#1{#1}\fi
\ifx\barticle  \undefined \def\barticle#1{}\fi
\ifx\binstitute  \undefined \def\binstitute#1{#1}\fi
\ifx\bpublisher  \undefined \def\bpublisher#1{#1}\fi
\ifx\doiurl    \undefined
  \def\doiurl#1{\href{http://dx.doi.org/#1}{\textsf{DOI}}}\fi
\ifx\arxivurl  \undefined
  \def\arxivurl#1{\href{http://arxiv.org/abs/#1}{\textsf{arXiv}}}\fi
\ifx\adsurl    \undefined
  \def\adsurl#1{\href{http://adsabs.harvard.edu/abs/#1}{\textsf{ADS}}}\fi
\ifx\botherref \undefined \def\botherref#1{}\fi
\ifx\url       \undefined \def\url#1{\textsf{#1}}\fi
\ifx\bchapter  \undefined \def\bchapter#1{}\fi
\ifx\bbook     \undefined \def\bbook#1{}\fi
\ifx\bcomment  \undefined \def\bcomment#1{#1}\fi
\ifx\oauthor   \undefined \def\oauthor#1{#1}\fi
\ifx\citeauthoryear \undefined\def \citeauthoryear#1{#1}\fi
\ifx\endbibitem\undefined \def\endbibitem{}\fi
\ifx\bconflocation  \undefined \def\bconflocation#1{#1} \fi

\bibitem[\protect\citeauthoryear{{Astropy Collaboration}
  \textit{et~al.}}{2013}]{Astropy2013}
\begin{barticle}
\bauthor{\bsnm{{Astropy Collaboration}}},
\bauthor{\bsnm{Robitaille}, \binits{T.P.}},
\bauthor{\bsnm{Tollerud}, \binits{E.J.}},
\bauthor{\bsnm{Greenfield}, \binits{P.}},
\bauthor{\bsnm{Droettboom}, \binits{M.}},
\bauthor{\bsnm{Bray}, \binits{E.}},
\bauthor{\bsnm{Aldcroft}, \binits{T.}},
\bauthor{\bsnm{Davis}, \binits{M.}},
\bauthor{\bsnm{Ginsburg}, \binits{A.}},
\bauthor{\bsnm{{Price-Whelan}}, \binits{A.M.}},
\bauthor{\bsnm{Kerzendorf}, \binits{W.E.}},
\bauthor{\bsnm{Conley}, \binits{A.}},
\bauthor{\bsnm{Crighton}, \binits{N.}},
\bauthor{\bsnm{Barbary}, \binits{K.}},
\bauthor{\bsnm{Muna}, \binits{D.}},
\bauthor{\bsnm{Ferguson}, \binits{H.}},
\bauthor{\bsnm{Grollier}, \binits{F.}},
\bauthor{\bsnm{Parikh}, \binits{M.M.}},
\bauthor{\bsnm{Nair}, \binits{P.H.}},
\bauthor{\bsnm{Unther}, \binits{H.M.}},
\bauthor{\bsnm{Deil}, \binits{C.}},
\bauthor{\bsnm{Woillez}, \binits{J.}},
\bauthor{\bsnm{Conseil}, \binits{S.}},
\bauthor{\bsnm{Kramer}, \binits{R.}},
\bauthor{\bsnm{Turner}, \binits{J.E.H.}},
\bauthor{\bsnm{Singer}, \binits{L.}},
\bauthor{\bsnm{Fox}, \binits{R.}},
\bauthor{\bsnm{Weaver}, \binits{B.A.}},
\bauthor{\bsnm{Zabalza}, \binits{V.}},
\bauthor{\bsnm{Edwards}, \binits{Z.I.}},
\bauthor{\bsnm{Azalee~Bostroem}, \binits{K.}},
\bauthor{\bsnm{Burke}, \binits{D.J.}},
\bauthor{\bsnm{Casey}, \binits{A.R.}},
\bauthor{\bsnm{Crawford}, \binits{S.M.}},
\bauthor{\bsnm{Dencheva}, \binits{N.}},
\bauthor{\bsnm{Ely}, \binits{J.}},
\bauthor{\bsnm{Jenness}, \binits{T.}},
\bauthor{\bsnm{Labrie}, \binits{K.}},
\bauthor{\bsnm{Lim}, \binits{P.L.}},
\bauthor{\bsnm{Pierfederici}, \binits{F.}},
\bauthor{\bsnm{Pontzen}, \binits{A.}},
\bauthor{\bsnm{Ptak}, \binits{A.}},
\bauthor{\bsnm{Refsdal}, \binits{B.}},
\bauthor{\bsnm{Servillat}, \binits{M.}},
\bauthor{\bsnm{Streicher}, \binits{O.}}:
\byear{2013},
\batitle{Astropy: {{A}} community {{Python}} package for astronomy}.
\bjtitle{\aap}
\bvolume{558},
\bfpage{A33}.
\doiurl{10.1051/0004-6361/201322068}.
\adsurl{2013A\%26A...558A..33A}.
\end{barticle}
\endbibitem

\bibitem[\protect\citeauthoryear{{Astropy Collaboration}
  \textit{et~al.}}{2018}]{Astropy2018}
\begin{barticle}
\bauthor{\bsnm{{Astropy Collaboration}}},
\bauthor{\bsnm{{Price-Whelan}}, \binits{A.M.}},
\bauthor{\bsnm{Sipőcz}, \binits{B.M.}},
\bauthor{\bsnm{Günther}, \binits{H.M.}},
\bauthor{\bsnm{Lim}, \binits{P.L.}},
\bauthor{\bsnm{Crawford}, \binits{S.M.}},
\bauthor{\bsnm{Conseil}, \binits{S.}},
\bauthor{\bsnm{Shupe}, \binits{D.L.}},
\bauthor{\bsnm{Craig}, \binits{M.W.}},
\bauthor{\bsnm{Dencheva}, \binits{N.}},
\bauthor{\bsnm{Ginsburg}, \binits{A.}},
\bauthor{\bsnm{VanderPlas}, \binits{J.T.}},
\bauthor{\bsnm{Bradley}, \binits{L.D.}},
\bauthor{\bsnm{{Pérez-Suárez}}, \binits{D.}},
\bauthor{\bsnm{{de Val-Borro}}, \binits{M.}},
\bauthor{\bsnm{Aldcroft}, \binits{T.L.}},
\bauthor{\bsnm{Cruz}, \binits{K.L.}},
\bauthor{\bsnm{Robitaille}, \binits{T.P.}},
\bauthor{\bsnm{Tollerud}, \binits{E.J.}},
\bauthor{\bsnm{Ardelean}, \binits{C.}},
\bauthor{\bsnm{Babej}, \binits{T.}},
\bauthor{\bsnm{Bach}, \binits{Y.P.}},
\bauthor{\bsnm{Bachetti}, \binits{M.}},
\bauthor{\bsnm{Bakanov}, \binits{A.V.}},
\bauthor{\bsnm{Bamford}, \binits{S.P.}},
\bauthor{\bsnm{Barentsen}, \binits{G.}},
\bauthor{\bsnm{Barmby}, \binits{P.}},
\bauthor{\bsnm{Baumbach}, \binits{A.}},
\bauthor{\bsnm{Berry}, \binits{K.L.}},
\bauthor{\bsnm{Biscani}, \binits{F.}},
\bauthor{\bsnm{Boquien}, \binits{M.}},
\bauthor{\bsnm{Bostroem}, \binits{K.A.}},
\bauthor{\bsnm{Bouma}, \binits{L.G.}},
\bauthor{\bsnm{Brammer}, \binits{G.B.}},
\bauthor{\bsnm{Bray}, \binits{E.M.}},
\bauthor{\bsnm{Breytenbach}, \binits{H.}},
\bauthor{\bsnm{Buddelmeijer}, \binits{H.}},
\bauthor{\bsnm{Burke}, \binits{D.J.}},
\bauthor{\bsnm{Calderone}, \binits{G.}},
\bauthor{\bsnm{Cano~Rodríguez}, \binits{J.L.}},
\bauthor{\bsnm{Cara}, \binits{M.}},
\bauthor{\bsnm{Cardoso}, \binits{J.V.M.}},
\bauthor{\bsnm{Cheedella}, \binits{S.}},
\bauthor{\bsnm{Copin}, \binits{Y.}},
\bauthor{\bsnm{Corrales}, \binits{L.}},
\bauthor{\bsnm{Crichton}, \binits{D.}},
\bauthor{\bsnm{D'Avella}, \binits{D.}},
\bauthor{\bsnm{Deil}, \binits{C.}},
\bauthor{\bsnm{Depagne}, \binits{{\'E}.}},
\bauthor{\bsnm{Dietrich}, \binits{J.P.}},
\bauthor{\bsnm{Donath}, \binits{A.}},
\bauthor{\bsnm{Droettboom}, \binits{M.}},
\bauthor{\bsnm{Earl}, \binits{N.}},
\bauthor{\bsnm{Erben}, \binits{T.}},
\bauthor{\bsnm{Fabbro}, \binits{S.}},
\bauthor{\bsnm{Ferreira}, \binits{L.A.}},
\bauthor{\bsnm{Finethy}, \binits{T.}},
\bauthor{\bsnm{Fox}, \binits{R.T.}},
\bauthor{\bsnm{Garrison}, \binits{L.H.}},
\bauthor{\bsnm{Gibbons}, \binits{S.L.J.}},
\bauthor{\bsnm{Goldstein}, \binits{D.A.}},
\bauthor{\bsnm{Gommers}, \binits{R.}},
\bauthor{\bsnm{Greco}, \binits{J.P.}},
\bauthor{\bsnm{Greenfield}, \binits{P.}},
\bauthor{\bsnm{Groener}, \binits{A.M.}},
\bauthor{\bsnm{Grollier}, \binits{F.}},
\bauthor{\bsnm{Hagen}, \binits{A.}},
\bauthor{\bsnm{Hirst}, \binits{P.}},
\bauthor{\bsnm{Homeier}, \binits{D.}},
\bauthor{\bsnm{Horton}, \binits{A.J.}},
\bauthor{\bsnm{Hosseinzadeh}, \binits{G.}},
\bauthor{\bsnm{Hu}, \binits{L.}},
\bauthor{\bsnm{Hunkeler}, \binits{J.S.}},
\bauthor{\bsnm{Ivezić}, \binits{{\v Z}.}},
\bauthor{\bsnm{Jain}, \binits{A.}},
\bauthor{\bsnm{Jenness}, \binits{T.}},
\bauthor{\bsnm{Kanarek}, \binits{G.}},
\bauthor{\bsnm{Kendrew}, \binits{S.}},
\bauthor{\bsnm{Kern}, \binits{N.S.}},
\bauthor{\bsnm{Kerzendorf}, \binits{W.E.}},
\bauthor{\bsnm{Khvalko}, \binits{A.}},
\bauthor{\bsnm{King}, \binits{J.}},
\bauthor{\bsnm{Kirkby}, \binits{D.}},
\bauthor{\bsnm{Kulkarni}, \binits{A.M.}},
\bauthor{\bsnm{Kumar}, \binits{A.}},
\bauthor{\bsnm{Lee}, \binits{A.}},
\bauthor{\bsnm{Lenz}, \binits{D.}},
\bauthor{\bsnm{Littlefair}, \binits{S.P.}},
\bauthor{\bsnm{Ma}, \binits{Z.}},
\bauthor{\bsnm{Macleod}, \binits{D.M.}},
\bauthor{\bsnm{Mastropietro}, \binits{M.}},
\bauthor{\bsnm{McCully}, \binits{C.}},
\bauthor{\bsnm{Montagnac}, \binits{S.}},
\bauthor{\bsnm{Morris}, \binits{B.M.}},
\bauthor{\bsnm{Mueller}, \binits{M.}},
\bauthor{\bsnm{Mumford}, \binits{S.J.}},
\bauthor{\bsnm{Muna}, \binits{D.}},
\bauthor{\bsnm{Murphy}, \binits{N.A.}},
\bauthor{\bsnm{Nelson}, \binits{S.}},
\bauthor{\bsnm{Nguyen}, \binits{G.H.}},
\bauthor{\bsnm{Ninan}, \binits{J.P.}},
\bauthor{\bsnm{Nöthe}, \binits{M.}},
\bauthor{\bsnm{Ogaz}, \binits{S.}},
\bauthor{\bsnm{Oh}, \binits{S.}},
\bauthor{\bsnm{Parejko}, \binits{J.K.}},
\bauthor{\bsnm{Parley}, \binits{N.}},
\bauthor{\bsnm{Pascual}, \binits{S.}},
\bauthor{\bsnm{Patil}, \binits{R.}},
\bauthor{\bsnm{Patil}, \binits{A.A.}},
\bauthor{\bsnm{Plunkett}, \binits{A.L.}},
\bauthor{\bsnm{Prochaska}, \binits{J.X.}},
\bauthor{\bsnm{Rastogi}, \binits{T.}},
\bauthor{\bsnm{Reddy~Janga}, \binits{V.}},
\bauthor{\bsnm{Sabater}, \binits{J.}},
\bauthor{\bsnm{Sakurikar}, \binits{P.}},
\bauthor{\bsnm{Seifert}, \binits{M.}},
\bauthor{\bsnm{Sherbert}, \binits{L.E.}},
\bauthor{\bsnm{{Sherwood-Taylor}}, \binits{H.}},
\bauthor{\bsnm{Shih}, \binits{A.Y.}},
\bauthor{\bsnm{Sick}, \binits{J.}},
\bauthor{\bsnm{Silbiger}, \binits{M.T.}},
\bauthor{\bsnm{Singanamalla}, \binits{S.}},
\bauthor{\bsnm{Singer}, \binits{L.P.}},
\bauthor{\bsnm{Sladen}, \binits{P.H.}},
\bauthor{\bsnm{Sooley}, \binits{K.A.}},
\bauthor{\bsnm{Sornarajah}, \binits{S.}},
\bauthor{\bsnm{Streicher}, \binits{O.}},
\bauthor{\bsnm{Teuben}, \binits{P.}},
\bauthor{\bsnm{Thomas}, \binits{S.W.}},
\bauthor{\bsnm{Tremblay}, \binits{G.R.}},
\bauthor{\bsnm{Turner}, \binits{J.E.H.}},
\bauthor{\bsnm{Terrón}, \binits{V.}},
\bauthor{\bsnm{{van Kerkwijk}}, \binits{M.H.}},
\bauthor{\bsnm{{de la Vega}}, \binits{A.}},
\bauthor{\bsnm{Watkins}, \binits{L.L.}},
\bauthor{\bsnm{Weaver}, \binits{B.A.}},
\bauthor{\bsnm{Whitmore}, \binits{J.B.}},
\bauthor{\bsnm{Woillez}, \binits{J.}},
\bauthor{\bsnm{Zabalza}, \binits{V.}},
\bauthor{\bsnm{{Astropy Contributors}}}:
\byear{2018},
\batitle{The {{Astropy Project}}: {{Building}} an {{Open}}-science {{Project}}
  and {{Status}} of the v2.0 {{Core Package}}}.
\bjtitle{\aj}
\bvolume{156},
\bfpage{123}.
\doiurl{10.3847/1538-3881/aabc4f}.
\adsurl{2018AJ....156..123A}.
\end{barticle}
\endbibitem

\bibitem[\protect\citeauthoryear{Auchère
  \textit{et~al.}}{2016}]{AuchereEtAl2016a}
\begin{barticle}
\bauthor{\bsnm{Auchère}, \binits{F.}},
\bauthor{\bsnm{Froment}, \binits{C.}},
\bauthor{\bsnm{Bocchialini}, \binits{K.}},
\bauthor{\bsnm{Buchlin}, \binits{E.}},
\bauthor{\bsnm{Solomon}, \binits{J.}}:
\byear{2016},
\batitle{On the {{Fourier}} and {{Wavelet Analysis}} of {{Coronal Time
  Series}}}.
\bjtitle{\apj}
\bvolume{825},
\bfpage{110}.
\doiurl{10.3847/0004-637X/825/2/110}.
\adsurl{2016ApJ...825..110A}.
\end{barticle}
\endbibitem

\bibitem[\protect\citeauthoryear{Boerner
  \textit{et~al.}}{2012}]{BoernerEtAl2012}
\begin{barticle}
\bauthor{\bsnm{Boerner}, \binits{P.}},
\bauthor{\bsnm{Edwards}, \binits{C.}},
\bauthor{\bsnm{Lemen}, \binits{J.}},
\bauthor{\bsnm{Rausch}, \binits{A.}},
\bauthor{\bsnm{Schrijver}, \binits{C.}},
\bauthor{\bsnm{Shine}, \binits{R.}},
\bauthor{\bsnm{Shing}, \binits{L.}},
\bauthor{\bsnm{Stern}, \binits{R.}},
\bauthor{\bsnm{Tarbell}, \binits{T.}},
\bauthor{\bsnm{Title}, \binits{A.}},
\bauthor{\bsnm{Wolfson}, \binits{C.J.}},
\bauthor{\bsnm{Soufli}, \binits{R.}},
\bauthor{\bsnm{Spiller}, \binits{E.}},
\bauthor{\bsnm{Gullikson}, \binits{E.}},
\bauthor{\bsnm{McKenzie}, \binits{D.}},
\bauthor{\bsnm{Windt}, \binits{D.}},
\bauthor{\bsnm{Golub}, \binits{L.}},
\bauthor{\bsnm{Podgorski}, \binits{W.}},
\bauthor{\bsnm{Testa}, \binits{P.}},
\bauthor{\bsnm{Weber}, \binits{M.}}:
\byear{2012},
\batitle{Initial {{Calibration}} of the {{Atmospheric Imaging Assembly}}
  ({{AIA}}) on the {{Solar Dynamics Observatory}} ({{SDO}})}.
\bjtitle{\solphys}
\bvolume{275}(\bissue{1-2}),
\bfpage{41}.
\doiurl{10.1007/s11207-011-9804-8}.
\adsurl{2012SoPh..275...41B}.
\end{barticle}
\endbibitem

\bibitem[\protect\citeauthoryear{Brooks, Warren, and
  {Ugarte-Urra}}{2012}]{BrooksEtAl2012}
\begin{barticle}
\bauthor{\bsnm{Brooks}, \binits{D.H.}},
\bauthor{\bsnm{Warren}, \binits{H.P.}},
\bauthor{\bsnm{{Ugarte-Urra}}, \binits{I.}}:
\byear{2012},
\batitle{Solar {{Coronal Loops Resolved}} by {{Hinode}} and the {{Solar
  Dynamics Observatory}}}.
\bjtitle{\apj}
\bvolume{755}(\bissue{2}),
\bfpage{L33}.
\doiurl{10.1088/2041-8205/755/2/L33}.
\adsurl{2012ApJ...755L..33B}.
\end{barticle}
\endbibitem

\bibitem[\protect\citeauthoryear{Culhane
  \textit{et~al.}}{2007}]{CulhaneEtAl2007}
\begin{barticle}
\bauthor{\bsnm{Culhane}, \binits{J.L.}},
\bauthor{\bsnm{Harra}, \binits{L.K.}},
\bauthor{\bsnm{James}, \binits{A.M.}},
\bauthor{\bsnm{{Al-Janabi}}, \binits{K.}},
\bauthor{\bsnm{Bradley}, \binits{L.J.}},
\bauthor{\bsnm{Chaudry}, \binits{R.A.}},
\bauthor{\bsnm{Rees}, \binits{K.}},
\bauthor{\bsnm{Tandy}, \binits{J.A.}},
\bauthor{\bsnm{Thomas}, \binits{P.}},
\bauthor{\bsnm{Whillock}, \binits{M.C.R.}},
\bauthor{\bsnm{Winter}, \binits{B.}},
\bauthor{\bsnm{Doschek}, \binits{G.A.}},
\bauthor{\bsnm{Korendyke}, \binits{C.M.}},
\bauthor{\bsnm{Brown}, \binits{C.M.}},
\bauthor{\bsnm{Myers}, \binits{S.}},
\bauthor{\bsnm{Mariska}, \binits{J.}},
\bauthor{\bsnm{Seely}, \binits{J.}},
\bauthor{\bsnm{Lang}, \binits{J.}},
\bauthor{\bsnm{Kent}, \binits{B.J.}},
\bauthor{\bsnm{Shaughnessy}, \binits{B.M.}},
\bauthor{\bsnm{Young}, \binits{P.R.}},
\bauthor{\bsnm{Simnett}, \binits{G.M.}},
\bauthor{\bsnm{Castelli}, \binits{C.M.}},
\bauthor{\bsnm{Mahmoud}, \binits{S.}},
\bauthor{\bsnm{{Mapson-Menard}}, \binits{H.}},
\bauthor{\bsnm{Probyn}, \binits{B.J.}},
\bauthor{\bsnm{Thomas}, \binits{R.J.}},
\bauthor{\bsnm{Davila}, \binits{J.}},
\bauthor{\bsnm{Dere}, \binits{K.}},
\bauthor{\bsnm{Windt}, \binits{D.}},
\bauthor{\bsnm{Shea}, \binits{J.}},
\bauthor{\bsnm{Hagood}, \binits{R.}},
\bauthor{\bsnm{Moye}, \binits{R.}},
\bauthor{\bsnm{Hara}, \binits{H.}},
\bauthor{\bsnm{Watanabe}, \binits{T.}},
\bauthor{\bsnm{Matsuzaki}, \binits{K.}},
\bauthor{\bsnm{Kosugi}, \binits{T.}},
\bauthor{\bsnm{Hansteen}, \binits{V.}},
\bauthor{\bsnm{Wikstol}, \binits{{\O}.}}:
\byear{2007},
\batitle{The {{EUV Imaging Spectrometer}} for {{Hinode}}}.
\bjtitle{\solphys}
\bvolume{243},
\bfpage{19}.
\doiurl{10.1007/s01007-007-0293-1}.
\adsurl{2007SoPh..243...19C}.
\end{barticle}
\endbibitem

\bibitem[\protect\citeauthoryear{Del~Zanna, O'Dwyer, and
  Mason}{2011}]{DelZannaEtAl2011_warm_loops}
\begin{barticle}
\bauthor{\bsnm{Del~Zanna}, \binits{G.}},
\bauthor{\bsnm{O'Dwyer}, \binits{B.}},
\bauthor{\bsnm{Mason}, \binits{H.E.}}:
\byear{2011},
\batitle{{{SDO AIA}} and {{Hinode EIS}} observations of "warm" loops}.
\bjtitle{\aap}
\bvolume{535},
\bfpage{A46}.
\doiurl{10.1051/0004-6361/201117470}.
\adsurl{2011A\&A...535A..46D}.
\end{barticle}
\endbibitem

\bibitem[\protect\citeauthoryear{Delaboudinière
  \textit{et~al.}}{1995}]{DelaboudiniereEtAl1995}
\begin{barticle}
\bauthor{\bsnm{Delaboudinière}, \binits{J.-P.}},
\bauthor{\bsnm{Artzner}, \binits{G.E.}},
\bauthor{\bsnm{Brunaud}, \binits{J.}},
\bauthor{\bsnm{Gabriel}, \binits{A.H.}},
\bauthor{\bsnm{Hochedez}, \binits{J.F.}},
\bauthor{\bsnm{Millier}, \binits{F.}},
\bauthor{\bsnm{Song}, \binits{X.Y.}},
\bauthor{\bsnm{Au}, \binits{B.}},
\bauthor{\bsnm{Dere}, \binits{K.P.}},
\bauthor{\bsnm{Howard}, \binits{R.A.}},
\bauthor{\bsnm{Kreplin}, \binits{R.}},
\bauthor{\bsnm{Michels}, \binits{D.J.}},
\bauthor{\bsnm{Moses}, \binits{J.D.}},
\bauthor{\bsnm{Defise}, \binits{J.M.}},
\bauthor{\bsnm{Jamar}, \binits{C.}},
\bauthor{\bsnm{Rochus}, \binits{P.}},
\bauthor{\bsnm{Chauvineau}, \binits{J.P.}},
\bauthor{\bsnm{Marioge}, \binits{J.P.}},
\bauthor{\bsnm{Catura}, \binits{R.C.}},
\bauthor{\bsnm{Lemen}, \binits{J.R.}},
\bauthor{\bsnm{Shing}, \binits{L.}},
\bauthor{\bsnm{Stern}, \binits{R.A.}},
\bauthor{\bsnm{Gurman}, \binits{J.B.}},
\bauthor{\bsnm{Neupert}, \binits{W.M.}},
\bauthor{\bsnm{Maucherat}, \binits{A.}},
\bauthor{\bsnm{Clette}, \binits{F.}},
\bauthor{\bsnm{Cugnon}, \binits{P.}},
\bauthor{\bsnm{{van Dessel}}, \binits{E.L.}}:
\byear{1995},
\batitle{{{EIT}}: {{Extreme}}-{{Ultraviolet Imaging Telescope}} for the {{SOHO
  Mission}}}.
\bjtitle{\solphys}
\bvolume{162},
\bfpage{291}.
\doiurl{10.1007/BF00733432}.
\adsurl{1995SoPh..162..291D}.
\end{barticle}
\endbibitem

\bibitem[\protect\citeauthoryear{Domingo, Fleck, and
  Poland}{1995}]{DomingoEtAl1995}
\begin{barticle}
\bauthor{\bsnm{Domingo}, \binits{V.}},
\bauthor{\bsnm{Fleck}, \binits{B.}},
\bauthor{\bsnm{Poland}, \binits{A.I.}}:
\byear{1995},
\batitle{The {{SOHO Mission}}: An {{Overview}}}.
\bjtitle{\solphys}
\bvolume{162},
\bfpage{1}.
\doiurl{10.1007/BF00733425}.
\adsurl{1995SoPh..162....1D}.
\end{barticle}
\endbibitem

\bibitem[\protect\citeauthoryear{Freeland and Handy}{2012}]{FreelandHandy2012}
\begin{botherref}
\oauthor{\bsnm{Freeland}, \binits{S.L.}},
\oauthor{\bsnm{Handy}, \binits{B.N.}}:
2012,
{{SolarSoft}}: {{Programming}} and data analysis environment for solar physics.
\textit{Astrophysics Source Code Library},
ascl:1208.013.
\adsurl{2012ascl.soft08013F}.
\end{botherref}
\endbibitem

\bibitem[\protect\citeauthoryear{Gabriel
  \textit{et~al.}}{2002}]{GabrielEtAl2002}
\begin{barticle}
\bauthor{\bsnm{Gabriel}, \binits{A.H.}},
\bauthor{\bsnm{Baudin}, \binits{F.}},
\bauthor{\bsnm{Boumier}, \binits{P.}},
\bauthor{\bsnm{García}, \binits{R.A.}},
\bauthor{\bsnm{{Turck-Chièze}}, \binits{S.}},
\bauthor{\bsnm{Appourchaux}, \binits{T.}},
\bauthor{\bsnm{Bertello}, \binits{L.}},
\bauthor{\bsnm{Berthomieu}, \binits{G.}},
\bauthor{\bsnm{Charra}, \binits{J.}},
\bauthor{\bsnm{Gough}, \binits{D.O.}},
\bauthor{\bsnm{Pallé}, \binits{P.L.}},
\bauthor{\bsnm{Provost}, \binits{J.}},
\bauthor{\bsnm{Renaud}, \binits{C.}},
\bauthor{\bsnm{Robillot}, \binits{J.-M.}},
\bauthor{\bsnm{Roca~Cortés}, \binits{T.}},
\bauthor{\bsnm{Thiery}, \binits{S.}},
\bauthor{\bsnm{Ulrich}, \binits{R.K.}}:
\byear{2002},
\batitle{A search for solar g modes in the {{GOLF}} data}.
\bjtitle{\aap}
\bvolume{390},
\bfpage{1119}.
\doiurl{10.1051/0004-6361:20020695}.
\adsurl{2002A\&A...390.1119G}.
\end{barticle}
\endbibitem

\bibitem[\protect\citeauthoryear{Golub \textit{et~al.}}{2007}]{GolubEtAl2007}
\begin{barticle}
\bauthor{\bsnm{Golub}, \binits{L.}},
\bauthor{\bsnm{Deluca}, \binits{E.}},
\bauthor{\bsnm{Austin}, \binits{G.}},
\bauthor{\bsnm{Bookbinder}, \binits{J.}},
\bauthor{\bsnm{Caldwell}, \binits{D.}},
\bauthor{\bsnm{Cheimets}, \binits{P.}},
\bauthor{\bsnm{Cirtain}, \binits{J.}},
\bauthor{\bsnm{Cosmo}, \binits{M.}},
\bauthor{\bsnm{Reid}, \binits{P.}},
\bauthor{\bsnm{Sette}, \binits{A.}},
\bauthor{\bsnm{Weber}, \binits{M.}},
\bauthor{\bsnm{Sakao}, \binits{T.}},
\bauthor{\bsnm{Kano}, \binits{R.}},
\bauthor{\bsnm{Shibasaki}, \binits{K.}},
\bauthor{\bsnm{Hara}, \binits{H.}},
\bauthor{\bsnm{Tsuneta}, \binits{S.}},
\bauthor{\bsnm{Kumagai}, \binits{K.}},
\bauthor{\bsnm{Tamura}, \binits{T.}},
\bauthor{\bsnm{Shimojo}, \binits{M.}},
\bauthor{\bsnm{McCracken}, \binits{J.}},
\bauthor{\bsnm{Carpenter}, \binits{J.}},
\bauthor{\bsnm{Haight}, \binits{H.}},
\bauthor{\bsnm{Siler}, \binits{R.}},
\bauthor{\bsnm{Wright}, \binits{E.}},
\bauthor{\bsnm{Tucker}, \binits{J.}},
\bauthor{\bsnm{Rutledge}, \binits{H.}},
\bauthor{\bsnm{Barbera}, \binits{M.}},
\bauthor{\bsnm{Peres}, \binits{G.}},
\bauthor{\bsnm{Varisco}, \binits{S.}}:
\byear{2007},
\batitle{The {{X}}-{{Ray Telescope}} ({{XRT}}) for the {{Hinode Mission}}}.
\bjtitle{\solphys}
\bvolume{243},
\bfpage{63}.
\doiurl{10.1007/s11207-007-0182-1}.
\adsurl{2007SoPh..243...63G}.
\end{barticle}
\endbibitem

\bibitem[\protect\citeauthoryear{{Hara}}{2008}]{Hara2008}
\begin{bchapter}
\bauthor{\bsnm{{Hara}}, \binits{H.}}:
\byear{2008},
\bctitle{{Overview of EIS Performance}}.
In: \beditor{\bsnm{{Matthews}}, \binits{S.A.}},
\beditor{\bsnm{{Davis}}, \binits{J.M.}},
\beditor{\bsnm{{Harra}}, \binits{L.K.}} (eds.)
\bbtitle{First Results From Hinode},
\bsertitle{Astronomical Society of the Pacific Conference Series}
\bseriesno{397},
\bfpage{11}.
\adsurl{2008ASPC..397...11H}.
\end{bchapter}
\endbibitem

\bibitem[\protect\citeauthoryear{Horne and Baliunas}{1986}]{HorneBaliunas1986}
\begin{barticle}
\bauthor{\bsnm{Horne}, \binits{J.H.}},
\bauthor{\bsnm{Baliunas}, \binits{S.L.}}:
\byear{1986},
\batitle{A {{Prescription}} for {{Period Analysis}} of {{Unevenly Sampled Time
  Series}}}.
\bjtitle{\apj}
\bvolume{302},
\bfpage{757}.
\doiurl{10.1086/164037}.
\adsurl{1986ApJ...302..757H}.
\end{barticle}
\endbibitem

\bibitem[\protect\citeauthoryear{Kosugi \textit{et~al.}}{2007}]{KosugiEtAl2007}
\begin{barticle}
\bauthor{\bsnm{Kosugi}, \binits{T.}},
\bauthor{\bsnm{Matsuzaki}, \binits{K.}},
\bauthor{\bsnm{Sakao}, \binits{T.}},
\bauthor{\bsnm{Shimizu}, \binits{T.}},
\bauthor{\bsnm{Sone}, \binits{Y.}},
\bauthor{\bsnm{Tachikawa}, \binits{S.}},
\bauthor{\bsnm{Hashimoto}, \binits{T.}},
\bauthor{\bsnm{Minesugi}, \binits{K.}},
\bauthor{\bsnm{Ohnishi}, \binits{A.}},
\bauthor{\bsnm{Yamada}, \binits{T.}},
\bauthor{\bsnm{Tsuneta}, \binits{S.}},
\bauthor{\bsnm{Hara}, \binits{H.}},
\bauthor{\bsnm{Ichimoto}, \binits{K.}},
\bauthor{\bsnm{Suematsu}, \binits{Y.}},
\bauthor{\bsnm{Shimojo}, \binits{M.}},
\bauthor{\bsnm{Watanabe}, \binits{T.}},
\bauthor{\bsnm{Shimada}, \binits{S.}},
\bauthor{\bsnm{Davis}, \binits{J.M.}},
\bauthor{\bsnm{Hill}, \binits{L.D.}},
\bauthor{\bsnm{Owens}, \binits{J.K.}},
\bauthor{\bsnm{Title}, \binits{A.M.}},
\bauthor{\bsnm{Culhane}, \binits{J.L.}},
\bauthor{\bsnm{Harra}, \binits{L.K.}},
\bauthor{\bsnm{Doschek}, \binits{G.A.}},
\bauthor{\bsnm{Golub}, \binits{L.}}:
\byear{2007},
\batitle{The {{Hinode}} ({{Solar}}-{{B}}) {{Mission}}: {{An Overview}}}.
\bjtitle{\solphys}
\bvolume{243},
\bfpage{3}.
\doiurl{10.1007/s11207-007-9014-6}.
\adsurl{2007SoPh..243....3K}.
\end{barticle}
\endbibitem

\bibitem[\protect\citeauthoryear{Lemen \textit{et~al.}}{2012}]{LemenEtAl2012}
\begin{barticle}
\bauthor{\bsnm{Lemen}, \binits{J.R.}},
\bauthor{\bsnm{Title}, \binits{A.M.}},
\bauthor{\bsnm{Akin}, \binits{D.J.}},
\bauthor{\bsnm{Boerner}, \binits{P.F.}},
\bauthor{\bsnm{Chou}, \binits{C.}},
\bauthor{\bsnm{Drake}, \binits{J.F.}},
\bauthor{\bsnm{Duncan}, \binits{D.W.}},
\bauthor{\bsnm{Edwards}, \binits{C.G.}},
\bauthor{\bsnm{Friedlaender}, \binits{F.M.}},
\bauthor{\bsnm{Heyman}, \binits{G.F.}},
\bauthor{\bsnm{Hurlburt}, \binits{N.E.}},
\bauthor{\bsnm{Katz}, \binits{N.L.}},
\bauthor{\bsnm{Kushner}, \binits{G.D.}},
\bauthor{\bsnm{Levay}, \binits{M.}},
\bauthor{\bsnm{Lindgren}, \binits{R.W.}},
\bauthor{\bsnm{Mathur}, \binits{D.P.}},
\bauthor{\bsnm{McFeaters}, \binits{E.L.}},
\bauthor{\bsnm{Mitchell}, \binits{S.}},
\bauthor{\bsnm{Rehse}, \binits{R.A.}},
\bauthor{\bsnm{Schrijver}, \binits{C.J.}},
\bauthor{\bsnm{Springer}, \binits{L.A.}},
\bauthor{\bsnm{Stern}, \binits{R.A.}},
\bauthor{\bsnm{Tarbell}, \binits{T.D.}},
\bauthor{\bsnm{Wuelser}, \binits{J.-P.}},
\bauthor{\bsnm{Wolfson}, \binits{C.J.}},
\bauthor{\bsnm{Yanari}, \binits{C.}},
\bauthor{\bsnm{Bookbinder}, \binits{J.A.}},
\bauthor{\bsnm{Cheimets}, \binits{P.N.}},
\bauthor{\bsnm{Caldwell}, \binits{D.}},
\bauthor{\bsnm{Deluca}, \binits{E.E.}},
\bauthor{\bsnm{Gates}, \binits{R.}},
\bauthor{\bsnm{Golub}, \binits{L.}},
\bauthor{\bsnm{Park}, \binits{S.}},
\bauthor{\bsnm{Podgorski}, \binits{W.A.}},
\bauthor{\bsnm{Bush}, \binits{R.I.}},
\bauthor{\bsnm{Scherrer}, \binits{P.H.}},
\bauthor{\bsnm{Gummin}, \binits{M.A.}},
\bauthor{\bsnm{Smith}, \binits{P.}},
\bauthor{\bsnm{Auker}, \binits{G.}},
\bauthor{\bsnm{Jerram}, \binits{P.}},
\bauthor{\bsnm{Pool}, \binits{P.}},
\bauthor{\bsnm{Soufli}, \binits{R.}},
\bauthor{\bsnm{Windt}, \binits{D.L.}},
\bauthor{\bsnm{Beardsley}, \binits{S.}},
\bauthor{\bsnm{Clapp}, \binits{M.}},
\bauthor{\bsnm{Lang}, \binits{J.}},
\bauthor{\bsnm{Waltham}, \binits{N.}}:
\byear{2012},
\batitle{The {{Atmospheric Imaging Assembly}} ({{AIA}}) on the {{Solar Dynamics
  Observatory}} ({{SDO}})}.
\bjtitle{\solphys}
\bvolume{275},
\bfpage{17}.
\doiurl{10.1007/s11207-011-9776-8}.
\adsurl{2012SoPh..275...17L}.
\end{barticle}
\endbibitem

\bibitem[\protect\citeauthoryear{Lomb}{1976}]{Lomb1976}
\begin{barticle}
\bauthor{\bsnm{Lomb}, \binits{N.R.}}:
\byear{1976},
\batitle{Least-{{Squares Frequency Analysis}} of {{Unequally Spaced Data}}}.
\bjtitle{\apss}
\bvolume{39}(\bissue{2}),
\bfpage{447}.
\doiurl{10.1007/BF00648343}.
\adsurl{1976Ap\&SS..39..447L}.
\end{barticle}
\endbibitem

\bibitem[\protect\citeauthoryear{Mariska}{2016}]{EISSWN20}
\begin{bchapter}
\bauthor{\bsnm{Mariska}, \binits{J.}}:
\byear{2016},
\bctitle{{{EIS Software Note No}}. 20: {{EIS}}/{{AIA Coalignment}}}.
In: \bbtitle{{{SolarSoft Documentation}}}.
\end{bchapter}
\endbibitem

\bibitem[\protect\citeauthoryear{Pesnell, Thompson, and
  Chamberlin}{2012}]{PesnellEtAl2012}
\begin{barticle}
\bauthor{\bsnm{Pesnell}, \binits{W.D.}},
\bauthor{\bsnm{Thompson}, \binits{B.J.}},
\bauthor{\bsnm{Chamberlin}, \binits{P.C.}}:
\byear{2012},
\batitle{The {{Solar Dynamics Observatory}} ({{SDO}})}.
\bjtitle{\solphys}
\bvolume{275},
\bfpage{3}.
\doiurl{10.1007/s11207-011-9841-3}.
\adsurl{2012SoPh..275....3P}.
\end{barticle}
\endbibitem

\bibitem[\protect\citeauthoryear{Scargle}{1982}]{Scargle1982}
\begin{barticle}
\bauthor{\bsnm{Scargle}, \binits{J.D.}}:
\byear{1982},
\batitle{Studies in astronomical time series analysis. {{II}}. {{Statistical}}
  aspects of spectral analysis of unevenly spaced data.}
\bjtitle{\apj}
\bvolume{263},
\bfpage{835}.
\doiurl{10.1086/160554}.
\adsurl{1982ApJ...263..835S}.
\end{barticle}
\endbibitem

\bibitem[\protect\citeauthoryear{Scherrer
  \textit{et~al.}}{2012}]{ScherrerEtAl2012}
\begin{barticle}
\bauthor{\bsnm{Scherrer}, \binits{P.H.}},
\bauthor{\bsnm{Schou}, \binits{J.}},
\bauthor{\bsnm{Bush}, \binits{R.I.}},
\bauthor{\bsnm{Kosovichev}, \binits{A.G.}},
\bauthor{\bsnm{Bogart}, \binits{R.S.}},
\bauthor{\bsnm{Hoeksema}, \binits{J.T.}},
\bauthor{\bsnm{Liu}, \binits{Y.}},
\bauthor{\bsnm{Duvall}, \binits{T.L.}},
\bauthor{\bsnm{Zhao}, \binits{J.}},
\bauthor{\bsnm{Title}, \binits{A.M.}},
\bauthor{\bsnm{Schrijver}, \binits{C.J.}},
\bauthor{\bsnm{Tarbell}, \binits{T.D.}},
\bauthor{\bsnm{Tomczyk}, \binits{S.}}:
\byear{2012},
\batitle{The {{Helioseismic}} and {{Magnetic Imager}} ({{HMI}})
  {{Investigation}} for the {{Solar Dynamics Observatory}} ({{SDO}})}.
\bjtitle{\solphys}
\bvolume{275},
\bfpage{207}.
\doiurl{10.1007/s11207-011-9834-2}.
\adsurl{2012SoPh..275..207S}.
\end{barticle}
\endbibitem

\bibitem[\protect\citeauthoryear{Shimizu
  \textit{et~al.}}{2007}]{ShimizuEtAl2007}
\begin{barticle}
\bauthor{\bsnm{Shimizu}, \binits{T.}},
\bauthor{\bsnm{Katsukawa}, \binits{Y.}},
\bauthor{\bsnm{Matsuzaki}, \binits{K.}},
\bauthor{\bsnm{Ichimoto}, \binits{K.}},
\bauthor{\bsnm{Kano}, \binits{R.}},
\bauthor{\bsnm{Deluca}, \binits{E.E.}},
\bauthor{\bsnm{Lundquist}, \binits{L.L.}},
\bauthor{\bsnm{Weber}, \binits{M.}},
\bauthor{\bsnm{Tarbell}, \binits{T.D.}},
\bauthor{\bsnm{Shine}, \binits{R.A.}},
\bauthor{\bsnm{Sôma}, \binits{M.}},
\bauthor{\bsnm{Tsuneta}, \binits{S.}},
\bauthor{\bsnm{Sakao}, \binits{T.}},
\bauthor{\bsnm{Minesugi}, \binits{K.}}:
\byear{2007},
\batitle{Hinode {{Calibration}} for {{Precise Image Co}}-{{Alignment}} between
  {{SOT}} and {{XRT}} (2006 {{November}}-2007 {{April}})}.
\bjtitle{\pasj}
\bvolume{59},
\bfpage{S845}.
\doiurl{10.1093/pasj/59.sp3.S845}.
\adsurl{2007PASJ...59S.845S}.
\end{barticle}
\endbibitem

\bibitem[\protect\citeauthoryear{Tsuneta
  \textit{et~al.}}{2008}]{TsunetaEtAl2008}
\begin{barticle}
\bauthor{\bsnm{Tsuneta}, \binits{S.}},
\bauthor{\bsnm{Ichimoto}, \binits{K.}},
\bauthor{\bsnm{Katsukawa}, \binits{Y.}},
\bauthor{\bsnm{Nagata}, \binits{S.}},
\bauthor{\bsnm{Otsubo}, \binits{M.}},
\bauthor{\bsnm{Shimizu}, \binits{T.}},
\bauthor{\bsnm{Suematsu}, \binits{Y.}},
\bauthor{\bsnm{Nakagiri}, \binits{M.}},
\bauthor{\bsnm{Noguchi}, \binits{M.}},
\bauthor{\bsnm{Tarbell}, \binits{T.}},
\bauthor{\bsnm{Title}, \binits{A.}},
\bauthor{\bsnm{Shine}, \binits{R.}},
\bauthor{\bsnm{Rosenberg}, \binits{W.}},
\bauthor{\bsnm{Hoffmann}, \binits{C.}},
\bauthor{\bsnm{Jurcevich}, \binits{B.}},
\bauthor{\bsnm{Kushner}, \binits{G.}},
\bauthor{\bsnm{Levay}, \binits{M.}},
\bauthor{\bsnm{Lites}, \binits{B.}},
\bauthor{\bsnm{Elmore}, \binits{D.}},
\bauthor{\bsnm{Matsushita}, \binits{T.}},
\bauthor{\bsnm{Kawaguchi}, \binits{N.}},
\bauthor{\bsnm{Saito}, \binits{H.}},
\bauthor{\bsnm{Mikami}, \binits{I.}},
\bauthor{\bsnm{Hill}, \binits{L.D.}},
\bauthor{\bsnm{Owens}, \binits{J.K.}}:
\byear{2008},
\batitle{The {{Solar Optical Telescope}} for the {{Hinode Mission}}: {{An
  Overview}}}.
\bjtitle{\solphys}
\bvolume{249},
\bfpage{167}.
\doiurl{10.1007/s11207-008-9174-z}.
\adsurl{2008SoPh..249..167T}.
\end{barticle}
\endbibitem

\bibitem[\protect\citeauthoryear{Yoshimura and
  McKenzie}{2015}]{YoshimuraMcKenzie2015}
\begin{barticle}
\bauthor{\bsnm{Yoshimura}, \binits{K.}},
\bauthor{\bsnm{McKenzie}, \binits{D.E.}}:
\byear{2015},
\batitle{Calibration of {{Hinode}}/{{XRT}} for {{Coalignment}}}.
\bjtitle{\solphys}
\bvolume{290}(\bissue{8}),
\bfpage{2355}.
\doiurl{10.1007/s11207-015-0746-4}.
\adsurl{2015SoPh..290.2355Y}.
\end{barticle}
\endbibitem

\bibitem[\protect\citeauthoryear{Young}{2010}]{EISSWN4}
\begin{bchapter}
\bauthor{\bsnm{Young}, \binits{P.}}:
\byear{2010},
\bctitle{{{EIS Software Note No}}. 4: {{The EIS Slit Tilts}}}.
In: \bbtitle{{{SolarSoft Documentation}}}.
\end{bchapter}
\endbibitem

\bibitem[\protect\citeauthoryear{Young \textit{et~al.}}{2013}]{YoungEtAl2013}
\begin{barticle}
\bauthor{\bsnm{Young}, \binits{P.R.}},
\bauthor{\bsnm{Doschek}, \binits{G.A.}},
\bauthor{\bsnm{Warren}, \binits{H.P.}},
\bauthor{\bsnm{Hara}, \binits{H.}}:
\byear{2013},
\batitle{Properties of a {{Solar Flare Kernel Observed}} by {{Hinode}} and
  {{SDO}}}.
\bjtitle{\apj}
\bvolume{766}(\bissue{2}),
\bfpage{127}.
\doiurl{10.1088/0004-637X/766/2/127}.
\adsurl{2013ApJ...766..127Y}.
\end{barticle}
\endbibitem

\end{thebibliography}

\end{article}
\end{document}